# Doping Dependence of the Critical Current and Irreversibility Field in $Y_{1-x}Ca_xBa_2Cu_3O_{7-\delta}$

Anita Semwal[1,2], N. M. Strickland[2], A. Bubendorfer[2], S. H. Naqib[1,2], Swee Kuan Goh[1,2] and G. V. M. Williams[1,2]

[1]*MacDiarmid Institute, Victoria University of Wellington, P.O. 600, Wellington, New Zealand.*
[2]*Industrial Research, P.O. 31310, Lower Hutt, New Zealand*

## Abstract

Critical current measurements were made on thin films of $YBa_2Cu_3O_{7-\delta}$ and $Y_{0.8}Ca_{0.2}Ba_2Cu_3O_{7-\delta}$ produced by sol-gel and pulsed laser deposition, respectively. Both samples have high critical currents at low temperatures and the critical current, $J_c$, is largest in the overdoped $Y_{0.8}Ca_{0.2}Ba_2Cu_3O_{7-\delta}$ thin film. Annealing studies show a large increase in $J_c$ with increasing hole concentration and suggests that the maximum $J_c$ should occur for overdoped samples. The very overdoped region of the superconducting phase diagram was probed by studying bulk ceramic $Y_{0.8}Ca_{0.2}Ba_2Cu_3O_{7-\delta}$ samples, where it was found that the irreversibility field systematically increases with increasing hole concentration when scaled by the reduced temperature ($\tau = (1-T/T_c)$).

## 1. Introduction

Maximizing the superconducting critical current is a major goal of applied superconductivity research. While much attention has been focused on $Bi_2Sr_2Ca_2Cu_3O_{10+\delta}$ wire technology, it is now becoming apparent that large superconducting critical current densities, $J_c$, can be achieved in $YBa_2Cu_3O_{7-\delta}$ thin-film coated conductors (>200 A/cm$^{-1}$-width at 77K [1]). The high superconducting critical currents are clearly due to an improvement in the type and distribution of pinning centres as well as epitaxy and the stronger c-axis coupling. Some studies have also shown that the oxygen content and hole concentration are important parameters for achieving the highest critical currents in the high temperature superconducting cuprates (HTSC's) [2-5]. In one study it was shown that $J_c$ and the irreversibility temperature from a c-axis aligned $Y_{0.8}Ca_{0.2}Ba_2Cu_3O_{7-\delta}$ powder sample reaches a peak in the overdoped side of the superconducting phase diagram and near a hole concentration, $p$, of ~ 0.19 holes per Cu [2,3]. It has been argued that $p$ ~0.19 is the location of a quantum critical point [3] where the normal-state pseudogap closes [6] and the superconducting condensate density is maximized [2,3,7]. It is therefore of technical interest to investigate the hole concentration dependence of $J_c$ in $Y_{1-x}Ca_xBa_2Cu_3O_{7-\delta}$.

In this paper we report the results from measurements of $J_c$ on $YBa_2Cu_3O_{7-\delta}$ and $Y_{0.8}Ca_{0.2}Ba_2Cu_3O_{7-\delta}$ thin films. The sample without Ca has a superconducting transition temperature, $T_c$, of greater than 90 K, but it can only be moderately overdoped ($p \leq 0.187$). However, $Y_{0.8}Ca_{0.2}Ba_2Cu_3O_{7-\delta}$ can be oxygen loaded to a higher hole concentration, although the hole concentrations achieved in $Y_{0.8}Ca_{0.2}Ba_2Cu_3O_{7-\delta}$ thin films are significantly less than that from bulk $Y_{0.8}Ca_{0.2}Ba_2Cu_3O_{7-\delta}$. We have also measured the irreversibility field in bulk $Y_{0.8}Ca_{0.2}Ba_2Cu_3O_{7-\delta}$.

## 2. Experimental details

A $YBa_2Cu_3O_{7-\delta}$ thin film was prepared on a $SrTiO_3$ substrate using the sol-gel method described in detail elsewhere [8]. It was prepared by thermal decomposition of a spin-coated sol-gel precursor. The precursor is a simple metal-organic structure, based on a mixed metal complex of yttrium, barium and copper with malic acid, formed through ligand exchange with acetate starting materials. The malate complex is then further reacted with glycerol to form a polyester incorporating the $YBa_2Cu_3O_{7-\delta}$ cations. Malic acid is both an excellent metal binder, keeping all three species soluble in solution, and also promotes polymer formation. An intimate mixture of the cations is produced, which is shown thermogravimetrically to break down in a single step, depositing the cations together. This avoids the formation of crystallites of any suboxide species in the film. The $YBa_2Cu_3O_{7-\delta}$ phase was formed by heating for 2 hours at 770 °C with an oxygen partial pressure of $4\times10^{-4}$ bar. Oxygen loading occurred during the cool-down in a 100 % oxygen atmosphere. X-ray diffraction measurements confirmed that the $YBa_2Cu_3O_{7-\delta}$ film was orientated with the c-axis perpendicular to the substrate surface. The film thickness was estimated from scanning electron microscope measurements to be 230 nm.

The $Y_{0.8}Ca_{0.2}Ba_2Cu_3O_{7-\delta}$, thin film was also prepared on a $SrTiO_3$ substrate by pulsed laser deposition (PLD) using a *Lambda Physik* LPX210 KrF laser with a wavelength of 248 nm, as described in detail elsewhere [9]. The films were grown at 800 °C and a pressure of less than 1.2 mbar. The film thickness was estimated from atomic force microscope measurements to be 250 nm. X-ray diffraction measurements indicated that the $Y_{0.8}Ca_{0.2}Ba_2Cu_3O_{7-\delta}$ film was orientated with the c-axis perpendicular to the substrate surface.

A $Y_{0.8}Ca_{0.2}Ba_2Cu_3O_{7-\delta}$ bulk ceramic sample was made from the respective nitrates and oxides as described elsewhere [10]. The final sintering was at 950 °C in oxygen for 16 hours.

The sample was then removed from the hot zone of the furnace but still within the flowing oxygen. It was then reinserted when the furnace had reached 600 °C. The sample was held at this temperature for 12 hours and then ramped at a rate of 0.2 °C/minute to 350 °C, followed by an anneal for 24 hours at 350 °C and ramped at 2 °C/minute to room temperature. This annealing regime is necessary to achieve maximum oxygentaion and hence maximum overdoping.

All the samples were annealed at various temperatures and oxygen partial pressures to alter the oxygen content and hence the hole concentration in the $CuO_2$ planes. Unfortunately, the absolute oxygen content is difficult to measure in thin films. However, it is possible to indirectly infer changes in the oxygen content from the c-axis lattice parameter, because the c-axis lattice parameter is known to systematically increase with increasing $\delta$ from measurements on bulk $YBa_2Cu_3O_{7-\delta}$ [11]. For this reason, the c-axis lattice parameter was measured from the x-ray diffraction data to determine if each annealing leads to an increase or decrease in $\delta$.

The superconducting transition temperature and $J_c$ were measured using a vibrating sample magnetometer. An applied magnetic field of 0.5 mT was used for measuring $T_c$ and $J_c$ was measured for magnetic fields of up to 1 T and with the applied magnetic field perpendicular to the surface of the film. Irreversibility measurements were made on the bulk ceramic $Y_{0.8}Ca_{0.2}Ba_2Cu_3O_{7-\delta}$ sample using a SQUID magnetometer and applied magnetic fields of up to 6 T by performing zero-field-cooled and field-cooled measurements.

## 3. Results and Discussion

The magnetic field dependence of $J_c$ from the as-made $YBa_2Cu_3O_{7-\delta}$ thin film ($T_c$=90.8 K) can be seen in figure 1 for various temperatures, $T$. It was calculated from the magnetization loops using the method of Angadi *et al.* [12]. In this case $J_c$ is given by,

$$J_C = \frac{3\pi}{2\Theta t} \frac{m_\infty}{(dm/dH)} \quad (1)$$

where $t$ is the thickness of the film, $m_\infty$ is the difference in the moment value at a given field $H_x$ in the reverse and forward cycles of the loop, $dm/dH$ represents the slope of the reverse leg of the hysteresis loop at the field $H_x$ and $\Theta$ is given by $\ln(8R/t - \frac{1}{2})$ where R represents the radius of the supercurrent loop and it is calculated from the initial slope of the loop using the equation $dm/dH = -\pi^2 R^3/\Theta$.

It can be seen in figure 1 that the field-dependence of $J_c$ follows a power law, which indicates a weak-link regime for magnetic fields between 0.05 T and 1 T [13]. The data was fitted to $J_c = (B/B_0)^{-n1}$ and the results are shown by the solid curves in figure 1. We find that the exponent is 0.51 for $T \leq 61$ K and it has increased to 0.58 for $T = 72$ K. The coefficient for temperatures approaching $T_c$ is close to that reported from studies on bi-crystal $YBa_2Cu_3O_{7-\delta}$ thin films [14]. However, the smaller exponent below ~61 K indicates a weaker dependence of $J_c$ on the applied magnetic field at low temperatures.

The temperature-dependence of $J_c$ can be seen in figure 2 for the $YBa_2Cu_3O_{7-\delta}$ film that was annealed to change the oxygen content. $J_c$ was calculated from the magnetization loops using the method of Brandt and Indenbom [15]. X-ray diffraction measurements showed that the sample with a $T_c$ of 92.5 K had a larger c-axis and hence $\delta$ than the as-made sample, where the increase in $T_c$ indicates that the as-made sample is overdoped ($p > 0.16$). Studies on other $YBa_2Cu_3O_{7-\delta}$ thin films have found that $J_c$ can be fitted to a power law temperature-dependence. For this reason, $J_c$ was fitted to $J_c = J_{c,0}(1-T/T_c)^{n2}$ and the resultant fitted curves are plotted in figure 2 (solid curves). The as-made sample shows a departure from this relation for reduced temperatures, $\tau = (1 - T/T_c)$, of less than ~0.5 and hence the data were only fitted above this. The resultant exponent, $n2$, is 1.05 for $T_c$ values of 90.8 K and 92.5 K. However, in the underdoped region there is an increase in $n2$ to 1.52 for $T_c$=86K and 1.65 for $T_c$=73.9 K. A crossover from quasilinear to a concave temperature-dependence has also been reported from measurements on a $YBa_2Cu_3O_{7-\delta}$ thin film made by pulsed laser deposition [4]. It has been suggested that this crossover arises from a change in the nature and distribution of the intergrain weak-links in the underdoped region [16].

The effect of hole concentration on $J_c$ can be seen in figure 3 (right axis), where $J_c$ is plotted at 15 K (filled circles) and $\tau$=0.2 (crosses) as a function of hole concentration. The hole concentration was estimated from $T_c$ using the empirical $T_c(p)$ found in the HTSC's [17]. It is written as $T_c(p) = T_{c,Max}[1 - 82.6(p - 0.16)^2]$ and is plotted in figure 3 (solid curve and left axis). There is a clear increase in $J_c$ with increasing $p$ and it is largest for overdoped $YBa_2Cu_3O_{7-\delta}$. The increase in the low-temperature $J_c$ may arise from a corresponding increase in $U(0)\xi_{ab}(0)$ [2,5], where $U(0)$ is the condensate energy and $\xi_{ab}(0)$ is the superconducting coherence length at T = 0K. It has been shown that if $J_c$ is dominated at low temperatures by pinning then it is proportional to $U(0)\xi_{ab}(0)$ [2,5] and from studies on $Y_{0.8}Ca_{0.2}Ba_2Cu_3O_{7-\delta}$, $U(0)\xi_{ab}(0)$ maximizes in the overdoped region at $p$ ~0.2 [5].

It is not possible to probe the very overdoped region of the superconducting phase diagram ($p$ >0.19) using $YBa_2Cu_3O_{7-\delta}$ and hence we also studied a $Y_{0.8}Ca_{0.2}Ba_2Cu_3O_{7-\delta}$ thin film where hole-doping by Ca extends the doping range. The resultant temperature-dependence of $J_c$ can be seen in figure 4. Oxygen annealing the as-made sample decreases $T_c$ and hence increases the hole-concentration. The low temperature $J_c$ reaches a maximum of ~13 MA/cm$^2$, which is greater than that observed in the $YBa_2Cu_3O_{7-\delta}$ thin film, and it is essentially the same for oxygen annealing at 350 °C, 600 °C and 650 °C followed by cooling at 2°C/minute to room temperature. We suspect that the relatively fast cooling rate is still insufficiently fast to avoid oxygenation on the cooldown, with the result that there is little difference in $T_c$ and $J_c$. Doping may be better achieved by quenching or sealing in quartz tubes with an appropriate oxygen getter. The minimum $T_c$ in the overdoped region (74.5 K) is significantly greater than that achievable in bulk ceramic $Y_{0.8}Ca_{0.2}Ba_2Cu_3O_{7-\delta}$ (45 K [10]) processed using a final high temperature oxygen anneal followed by oxygen loading at a lower temperature. It is not known why this additional processing step leads to a significant increase in the hole concentration.

$J_c$ was also fitted to $J_c = J_{c,0}(1 - T/T_c)^{n2}$ (solid curves) and the resultant $n2$ is 2.54, 2.16, 1.96 and 1.77 for the as-made sample, and after annealing at 350 °C, 600 °C and 650 °C, respectively. These values are larger than those observed in the $YBa_2Cu_3O_{7-\delta}$ thin film in the optimally and lightly overdoped region. The concave temperature-dependence may suggest that a further increase in $J_c$ could be possible if the effect of the weak-links is reduced, leading to a value of $n2$ closer to that observed in fully loaded $YBa_2Cu_3O_{7-\delta}$. We note that significant changes can be seen in $J_c$ for $\tau$<0.6 where $J_c$ systematically increases with increasing annealing temperature. Since the associated changes in $T_c$ are small, it is possible that these changes may be due to a decrease in the weak-link behaviour.

We show in figure 3 the low-temperature $J_c$ against hole concentration for the $Y_{0.8}Ca_{0.2}Ba_2Cu_3O_{7-\delta}$ thin film (open circles). To enable a comparison with the $YBa_2Cu_3O_{7-\delta}$ thin film, the $J_c$ values are scaled by a factor of 0.54. The plot suggests that $J_c$ is largest in the overdoped region where $U(0)\xi_{ab}(0)$ is maximized.

As mentioned above, the very overdoped region cannot be accessed with $Y_{0.8}Ca_{0.2}Ba_2Cu_3O_{7-\delta}$ thin films. Therefore, to probe the very overdoped region, we performed irreversibility measurements on a bulk $Y_{0.8}Ca_{0.2}Ba_2Cu_3O_{7-\delta}$ ceramic sample. While there are a number of interpretations of the irreversibility field, $B_{irr}$, it is believed that large irreversibility fields are correlated with large critical currents. $B_{irr}$ can be written as $B_{irr} = B_{irr}(0)[1 - T/T_c]^{n3}$ for temperatures near $T_c$ where $n3=1.5$ for flux creep or flux-line melting [18,19] and $n3=2.0$ for a 3D to 2D flux-line transition [19]. However, these exponents can be larger when a temperature-dependent penetration depth anisotropy is included [20].

The irreversibility field from the bulk $Y_{0.8}Ca_{0.2}Ba_2Cu_3O_{7-\delta}$ ceramic sample is plotted in figure 5 for the as-made sample (filled circles, $T_c$=52 K, $p$=0.231) and then after annealing at different temperatures and oxygen partial pressures. Each annealing was followed by quenching in liquid nitrogen. The data in figure 5 were fitted to $B_{irr} = B_{irr}(0)[1 - T/T_c]^{n3}$ (solid curves) and the resultant values of ($B_{irr}(0), n3$) are (105 T, 1.85), (43 T, 2.08), (31 T, 2.41) and (12 T, 2.60) for $T_c$=52 K, 81 K, 85 K and 73 K respectively. The last sample is underdoped. The value of $n3$ from the very overdoped sample with $T_c$=52 K ($p$=0.231) is slightly larger than that reported in a study on c-axis aligned and sintered ceramic $YBa_2Cu_3O_{7-\delta}$ ($n3$=1.7 [5]) samples over a similar reduced temperature range and with $\delta \leq 0.2$. We note that it has been reported that samples with larger values of δ are better described by a 3D to 2D flux-line transition and hence $n3$ is expected to be larger for more underdoped $YBa_2Cu_3O_{7-\delta}$ [21].

Our measurements on bulk $Y_{0.8}Ca_{0.2}Ba_2Cu_3O_{7-\delta}$ show a clear increase in $B_{irr}(0)$ that is still increasing even for hole concentrations of up to 0.231. An increase in $B_{irr}(0)$ was also observed in $Bi_{2-x}Pb_xSr_2Ca_2Cu_3O_{10+\delta}$ ceramic samples up to the maximum achievable hole concentration ($p$~0.17) [20]. It was predicted that $B_{irr}(0)$ in $Bi_2Sr_2CaCu_3O_8$ and $Bi_{2-x}Pb_xSr_2Ca_2Cu_3O_{10+\delta}$ maximizes at $p$~0.21 [20]. This was based on the competing roles of $\lambda_{ab}^{-2}(0)$, which maximises at p ~ 0.19 and $\lambda_c^{-2}(0)$, which increases monotonically with doping across the entire phase diagram. The current study on $Y_{0.8}Ca_{0.2}Ba_2Cu_3O_{7-\delta}$ shows that $B_{irr}(0)$ is still increasing for hole concentrations greater than 0.21 and this presumably reflects the stronger doping dependence of $\lambda_c^{-2}$ in this compound.

The irreversibility temperature at a fixed field is plotted in the inset to figure 5 from the bulk $Y_{0.8}Ca_{0.2}Ba_2Cu_3O_{7-\delta}$ ceramic sample at 4 T. Data are also included for a sample annealed at 600 °C and then slowly cooled to room temperature (overdoped and $T_c$=70 K). As noted in a previous study on a c-axis aligned powder sample in resin [3], the irreversibility temperature maximizes in the overdoped region.

## 4. Conclusion

Critical current measurements on a $YBa_2Cu_3O_{7-\delta}$ thin film made using a sol-gel method displays a field-dependence for the c-axis parallel to the applied magnetic field and temperatures near $T_c$ that is similar to that seen in other $YBa_2Cu_3O_{7-\delta}$ thin films. For temperatures far below $T_c$, the critical current is a more slowly varying function of applied magnetic field for magnetic fields in the region of 0.05 T to 1 T. $J_c$ is also observed to increase with increasing hole concentration. The results suggest that the maximum $J_c$ should occur for fully-loaded and overdoped thin films.

Measurements of $J_c$ on a overdoped $Y_{0.8}Ca_{0.2}Ba_2Cu_3O_{7-\delta}$ thin film made by pulsed laser deposition show a higher $J_c$ at low temperatures. However, for this film $J_c$ for temperatures closer to $T_c$ is strongly dependent on the oxygen loading conditions. Irreversibility measurements on a more overdoped bulk $Y_{0.8}Ca_{0.2}Ba_2Cu_3O_{7-\delta}$ ceramic sample shows that $B_{irr}(0)$ is still increasing for hole concentrations as high as ~0.23. When the data are plotted as the irreversibility temperature at a fixed field, the irreversibility temperature is found to maximize in the overdoped region as noted in a previous study. The $J_c$ measurements on $Y_{1-x}Ca_xBa_2Cu_3O_{7-\delta}$ thin films indicate that $J_c$ is largest in the overdoped region where $U(0)\xi_{ab}(0)$ is maximized.

**Acknowledgements**
Authors thank Jeff Tallon for his suggestions. Funding support was provided by the MacDiarmid Institute for Advanced Materials and Nanotechnology, the New Zealand Marsden fund and the New Zealand Foundation for Research, Science and Technology.

**References**
[1] X. Li *et al.*, Physica C **390**, 249 (2003).
[2] J. L. Tallon, G. V. M. Williams and J. W. Loram, Physica C **238**, 9 (2000).
[3] J. L. Tallon, J. W. Loram, G. V. M. Williams, J. R. Cooper, I. R. Fisher, and C. Bernhard, Physica Status Solidi (b) **215**, 531 (1999).
[4] E. C. Jones, D. K. Christen, J. R. Thompson, R. Feenstra, S. Zhu, D. H. Lowndes, J. M. Phillips, M. P. Siegal and J. D. Budai, Phys. Rev. B **47**, 8986 (1993).
[5] J. G. Ossandon, J. R. Thompson, D. K. Christen, B. C. Sales, H. R. Kerchner, J. O. Thompson, Y. R. Sun, K. W. Lay, and J. E. Tkaczyk, Phys. Rev. B **45**, 12534 (1992).
[6] for a review see, G. V. M. Williams, *"NMR Studies of the Normal-State Pseudogap in High Temperature Superconducting Cuprates"*, in *"Studies of High Temperature Superconductors Vol 27"* (Nova Science Publishers, New York, 1999), pp113.
[7] J. L. Tallon, J. W. Loram, J. R. Cooper and C. Bernhard, Phys. Rev. B **68**, 180501 (2003).
[8] A. J. Bubendorfer, T. Kemmitt, L. J. Campbell and N. J. Long, IEEE Trans. Appl. Supercon. **13** , 2739 (2003).
[9] S. H. Naqib, R. A. Chakalov and J. R. Cooper, condmat/0312433 and appearing in Physica C (2004).
[10] J. L. Tallon, C. Bernhard, H. Shaked, R. L. Hitterrman and J. D. Jorgensen, Phys. Rev. B **51**, 12911 (1995).
[11] J. D. Jorgensen, B. W. Veal, A. P. Paulikas, L. J. Nowicki, G. W. Crabtree, H. Claus and W. K. Kwok,Phys. Rev. B **41**, 1863 (1990).
[12] M. A. Angadi, A. D. Caplin, J. R. Laverty Z. X. and Shen, Physica C **177**, 479 (1991).
[13] R. L. Petersen and J. W. Ekin, Phys. Rev. B **42**, 8014 (1990).
[14] B. Holzapfel, D. Verebelyi, C. Cantoni, M. Paranthaman, B. Sales, R. Feenstra, D. Christen and D. P. Norton, Physica C **341-348**, 1431 (2000).
[15] E. H. Brandt and M. Indenbom, Phys. Rev. B **48**, 12893 (1993).
[16] H. Darhmoaui and J. Jung, Phys. Rev. B **53**, 14621 (1996).
[17] M. R. Presland, J. L. Tallon, R. G. Buckley, R. S. Liu, and N. E. Flower, Physica C **176**, 95 (1991).
[18] Y. Yeshurun and A. P. Malozemoff, Phys. Rev. Lett. **60**, 2202 (1988).
[19] V. Hardy, Physica C **232**, 347 (1994).
[20] G. V. M. Williams, J. L. Tallon and D. M. Pooke, Phys. Rev. B **62**, 9132 (2000).
[21] J. H. Brewer *et al.*, Phys. Rev. B **61**, 890 (2000).

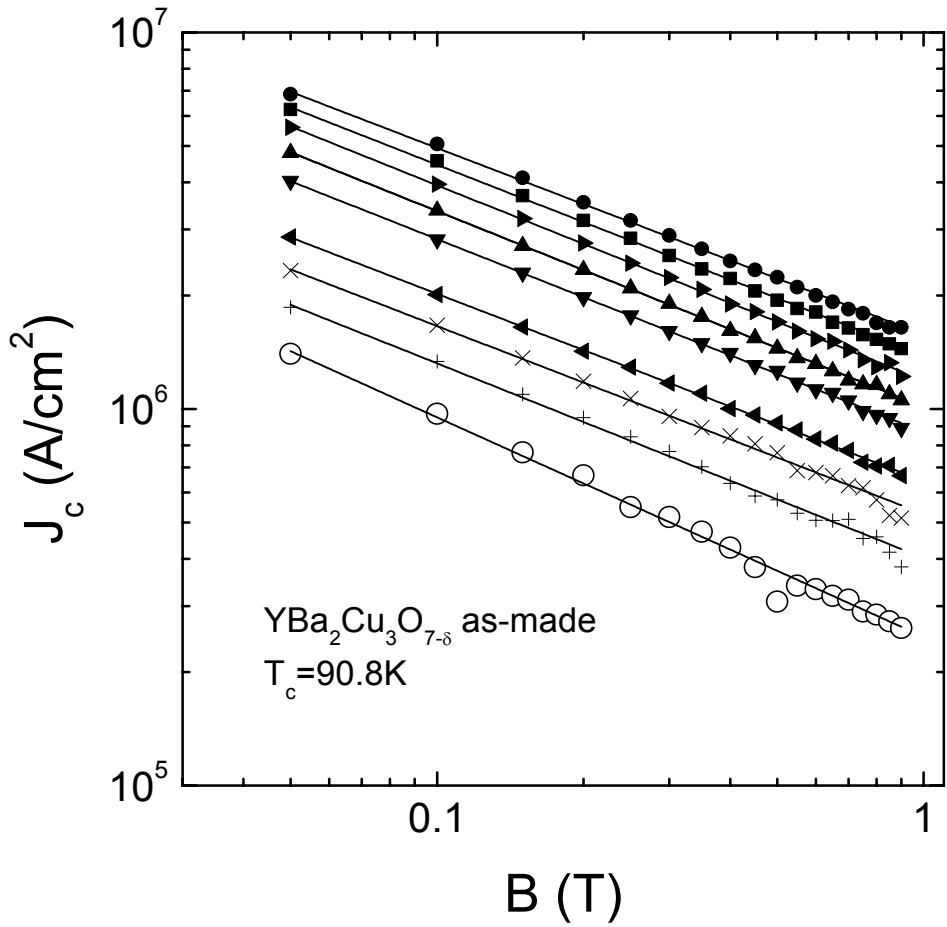

**Figure 1:** Plot of the magnetization $J_c$ against applied magnetic field for the as-made $YBa_2Cu_3O_{7-\delta}$ thin film for temperatures of 13 K (filled circles), 18 K (filled squares), 22 K (filled right triangles), 27 K (filled up triangles), 32 K (filled down triangles), 42 K (filled left triangles), 56 K (crosses), 61 K (plus symbols) and 72 K (open circles). The solid curves are fits to the data as described in the text.

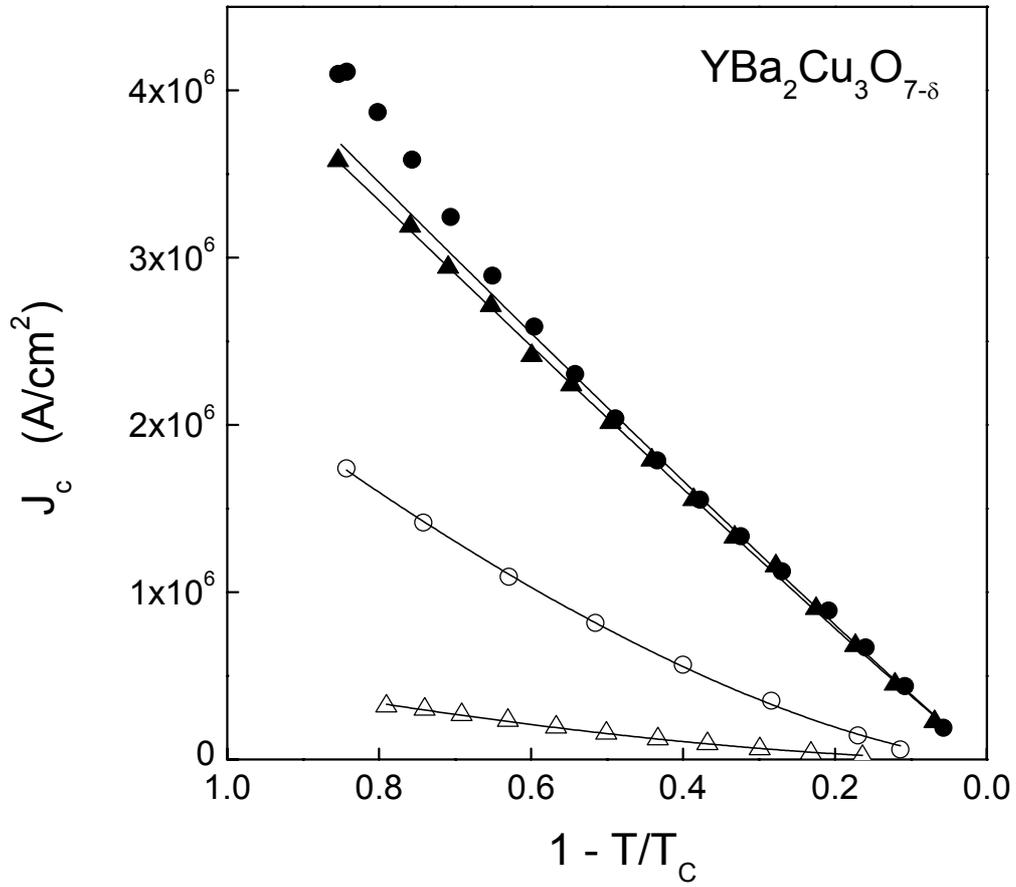

**Figure 2:** Plot of the magnetization $J_c$ against reduced temperature from a $YBa_2Cu_3O_{7-\delta}$ thin film and $T_c$ values of 90.8 K (filled circles), 92.5 K (filled up triangles), 86 K (open circles) and 73.9 K (open up triangles). The solid curves are fits to the data as described in the text.

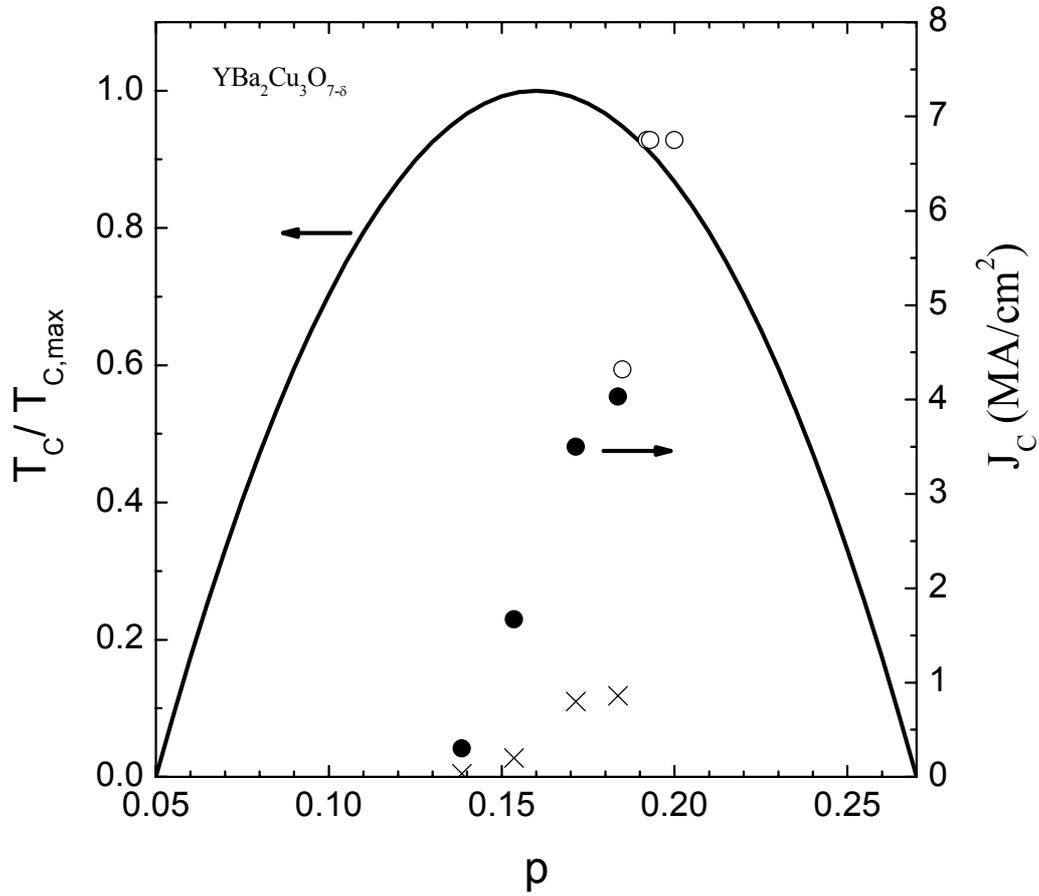

**Figure 3:** Plot of the magnetization $J_c$ against hole concentration (right axis) from a $YBa_2Cu_3O_{7-\delta}$ thin film at 15 K (solid circles) and $1-T/T_c=0.2$ (crosses) using the data plotted in figure 2. Also shown is the empirical $T_c/T_{c,Max}$ correlation found in the HTSC's (solid curve [15]). Data from the $Y_{0.8}Ca_{0.2}Ba_2Cu_3O_{7-\delta}$ thin film (see figure 4) at 15 K are plotted after scaling by a factor of 0.54 (open circles).

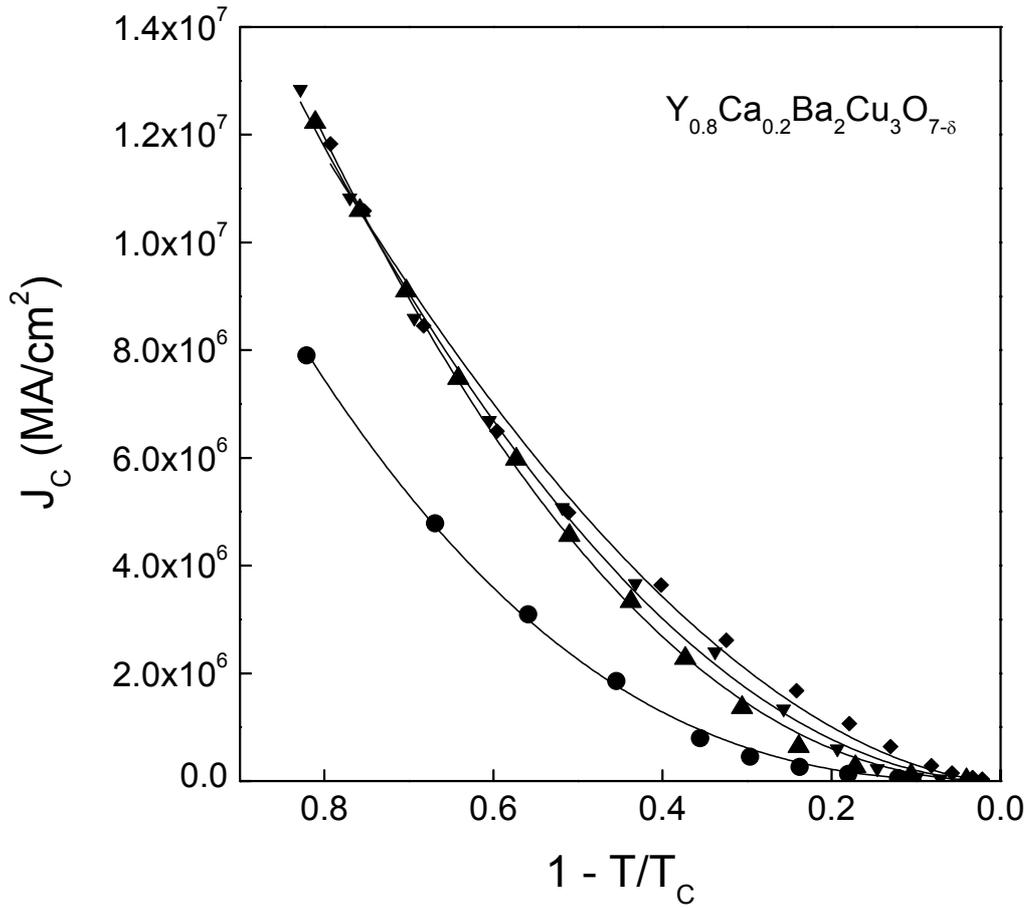

**Figure 4:** Plot of $J_c$ against temperature from a $Y_{0.8}Ca_{0.2}Ba_2Cu_3O_{7-\delta}$ thin film with $T_c$ values of 81.6 K (as-made, filled circles), 74.5 K (filled up triangles), 78.7 K (filled down triangles) and 78.2 K (filled diamonds). The annealing temperatures for the last three samples were 350 °C for four hours, 600 °C and 650 °C for two hours in oxygen. They were all cooled to room temperature at a rate of 2°C/minute. The solid curves are fits to the data as described in the text.

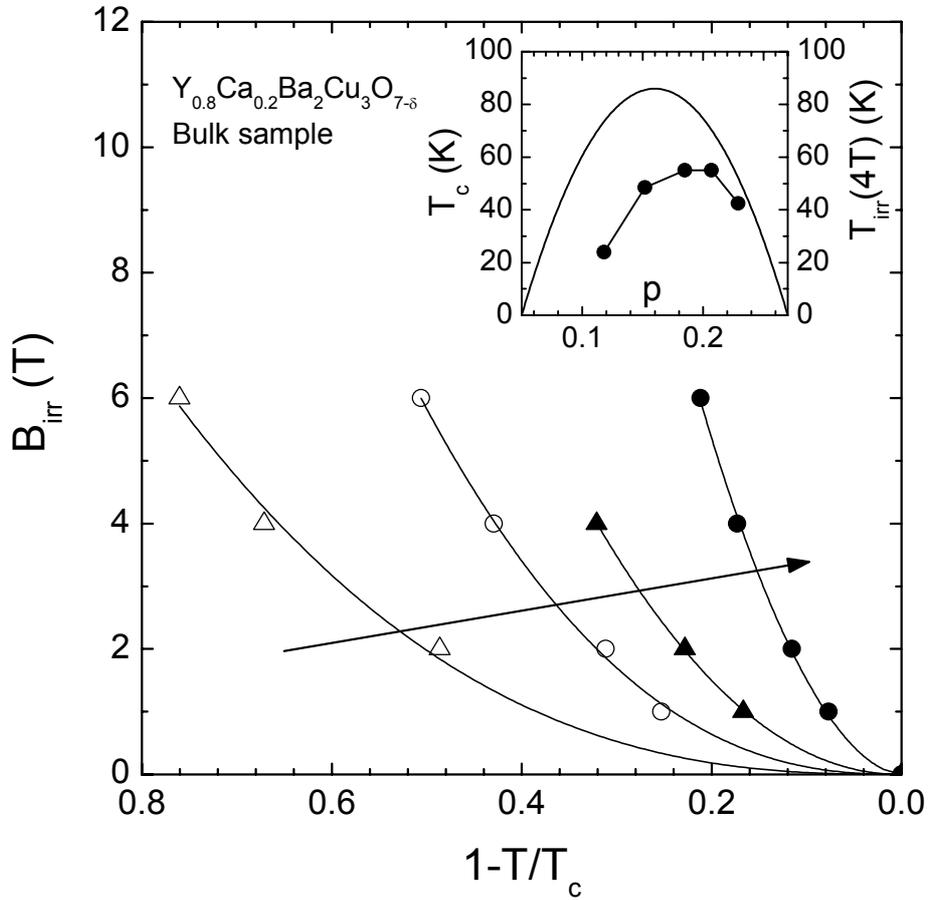

**Figure 5:** Plot of the irreversibility field, $B_{irr}$, against the reduced temperature from a bulk ceramic $Y_{0.8}Ca_{0.2}Ba_2Cu_3O_{7-\delta}$ sample with $T_c$ values of 52 K (overdoped, solid circles) and 81 K (overdoped, solid up triangles), 85 K (underdoped, open circles) and 73 K (underdoped, open up triangles). The arrow indicates increasing hole concentration. Inset: plot of the irreversibility temperature at 4 T (right axis and filled symbols) and the empirical $T_c$ scaling curve (left axis and solid curve [15]) against the hole concentration.